\documentclass[12pt]{article}
\usepackage[utf8x]{inputenc}
\usepackage{amsmath}
\usepackage{graphicx,psfrag,epsf}
\usepackage{enumerate}
\usepackage{natbib}
\usepackage{url}
\usepackage{longtable}
\usepackage{fancyhdr}
\newcommand{\blind}{1}

\begin{document}

\chead[C]{\scriptsize \textit{Draft under review in the Journal of the American Statistical Association. DO NOT REDISTRIBUTE}}
\fancyfoot[C]{\quad\quad\quad\quad\quad\quad\thepage}
\renewcommand{\headrulewidth}{0pt}
\renewcommand{\footrulewidth}{0pt}
\renewcommand{\sectionmark}[1]{}
\renewcommand{\subsectionmark}[1]{}

\def\spacingset#1{\renewcommand{\baselinestretch}%
{#1}\small\normalsize} \spacingset{1}


\if1\blind
{
  \title{\bf Data-driven causal path discovery without \\prior knowledge - a benchmark study}
  \author{Marcel Młyńczak\thanks{
    Marcel Młyńczak, PhD: Assistant Professor at Institute of Metrology and Biomedical Engineering, Faculty of Mechatronics, Warsaw University of Technology, Poland. }}
  \maketitle
  \thispagestyle{fancy} 
} \fi

\if0\blind
{
  \bigskip
  \bigskip
  \bigskip
  \begin{center}
    {\LARGE\bf Title}
\end{center}
  \medskip
} \fi

\bigskip
\begin{abstract}

Causal discovery broadens the inference possibilities, as correlation does not inform about the relationship direction. The common approaches were proposed for cases in which prior knowledge is desired, when the impact of a treatment/intervention variable is discovered or to analyze time-related dependencies. In some practical applications, more universal techniques are needed and have already been presented. Therefore, the aim of the study was to assess the accuracies in determining causal paths in a~dataset without considering the ground truth and the contextual information. This benchmark was performed on the database with cause-effect pairs, using a framework consisting of generalized correlations (GC), kernel regression gradients (GR) and absolute residuals criteria (AR), along with causal additive modeling (CAM). The best overall accuracy, $80\%$, was achieved for the (majority voting) combination of GC, AR, and CAM, however, the most similar sensitivity and specificity values were obtained for AR. Bootstrap simulation established the probability of correct causal path determination (which pairs should remain indeterminate). The mean accuracy was then improved to $83\%$ for the selected subset of pairs. The described approach can be used for preliminary dependence assessment, as an initial step for commonly used causality assessment frameworks or for comparison with prior assumptions.

\end{abstract}

\noindent%
\vfill

\newpage
\spacingset{1.45}

\section{Introduction}

The standard Pearson's correlation coefficient only provides information about the strength of a linear relationship between two variables. It cannot be used to establish the direction of the relation. However, in many economic and medical applications, the analysis and further inference should be extended with information about the cause-and-effect relationship between the variables  [\cite{Spirtes2010}].

Time series data are usually analyzed with the Granger causality framework or its generalizations and special extensions [\cite{Granger1969, Chen2004}]. The main concept is that a time-evolving variable X causes (in the Granger sense) another evolving variable Y if predictions of Y based only on its own past values are worse than those also considering past values of X. It is based on a meaningful and relatively obvious idea that the cause should happen before its effect. The improvement of prediction accuracy, in other words, the increase of the determination coefficient of a considered model, delivers specific and unique information that is propagated to future values of the observed effect variable. However, this methodology is restricted to time-related data.

Time-independent, data-frame-related techniques are also introduced. Many of them are based on graphical, structural, potential-outcome- or additive-noise-type approaches [\cite{Pearl2009,Pearl2010,Rubin1974,Mooij2016,Scholkopf2012}]. The first major school, tied to the Neyman–Rubin causal model framework, utilizes potential outcome analysis. The reasoning is that the result may depend on different possibilities/choices in the past, but only one has happened in reality. Initially, a randomized experiment should be performed on equivalent groups to try to attribute an effect based on several specific causes [\cite{Neyman1923}]. In many cases, such experiment cannot be carried out due to, e.g., long time frames, requiring too much interference in the lives of the study participants, or the possible unrecognizable influence of secondary variables. Therefore, Rubin et al. proposed a non-random assignment mechanism in which any considered secondary variables are balanced within the groups created by analyzed variables. The relation is then attributed by finding the difference between the effects on various interventions/treatments [\cite{Rubin2005}].  

Another school is connected with Pearl's research on causality, which primarily uses structural equation modeling (SEM) and its generalization from linear to nonparametric models. He proposed a way of interpreting the equations from a causal and counterfactual perspective [\cite{Pearl2009a}]. Furthermore, causal relations between the events can be established as a directed acyclic graph (DAG), which is a Bayesian network - each node can represent a state and each directed link has a probability measure, which describes the possibility of a transition. For such DAGs, the do-calculus rules theorem was introduced to analyze the effects (including counterfactual outcomes) of an intervention, e.g., \textit{do(x)}, where x is treated as an atomic intervention [\cite{Pearl1995}]. However, the presented approach desires prior knowledge: primarily, a preliminary directed acyclic graph may improve search process [\cite{Spirtes2000}]. Such causal discovery search comprises many methods, e.g., fast causal inference, fast greedy equivalance and Bayesian network estimators [\cite{Chickering2002,Zhang2006,Rebane2013,Zhang2017}].

On the other hand, it appears that in many physiological studies there is a search for a more universal technique. Firstly, the basic medical knowledge cannot always identify the cause, because the network of relationships might be bidirectional. Secondly, several hypotheses may not be strictly related with interventional, but rather with general descriptive variables, and they may try to describe generic, static relationships, e.g., between two systems. In such situations, it could be relevant to discover the causal path in a data-driven manner and only then relate the result to medical expectations. Using Bayesian-style approaches might require much more data for the correct transition from prior to posterior distribution and to produce proper inference, and that is not possible in every study. Moreover, some applications do not apply treatment variables and counterfactual effects.

Several approaches that fulfill these criteria were found. We finally chose the frameworks elaborated by Vinod et al., comprising generalized correlations, kernel regressions and stochastic dominance criteria, along with an algorithm for fitting a causal additive model (CAM) [\cite{Zheng2012,Buhlmann2014,Vinod2017}].

Therefore, the aim of the study was to benchmark these methods with the data from the cause-effect pairs database [\cite{Mooij2016,Dua2017}] (available also from a causality contest, the Cause-Effect Pairs Challenge, organized by Kaggle [\cite{CCC}]), by assessing the possibility of discovering a causal link with each method separately and for several combinations, without stating prior knowledge on the subject of the specific phenomenon.

\section{Materials \& Methods}

A set of 108 files, each consisting of variables (usually two) describing various physical, meteorological and economic cases, was shared. Pairs were removed from the list if the single-input data were not in the form of a single vector, if no ground truth about the causal variable was given or if the CAM algorithm reported a calculation error. Thus, 95 pairs were used for the benchmark study. The list of index numbers of the files taken into consideration, along with short descriptions of every case and other information and calculations, are provided in the supplementary materials (\textbf{T1}).

Four methods enabling time-invariant causal path discovery were elaborated; their short descriptions are given below:

\begin{itemize}
	\item (\textbf{M1}), a criterion based on minimizing local kernel regression gradients: if path $ X \rightarrow Y $ has smaller partial derivatives than the inverse model, it suggests the presented direction; implemented by equation (1) below [\cite{Vinod2017}];
	\item (\textbf{M2}), a criterion based on absolute values of residuals: if path $ X \rightarrow Y $ has smaller absolute residuals than the inverse model, it suggests the presented direction; implemented by equation (2) below [\cite{Vinod2017}];
	\item (\textbf{M3}), the asymmetric generalized correlations $ { r }_{ x|y }^{ * } $ and $ { r }_{ y|x }^{ * } $; if $ |{ r }_{ x|y }^{ * }| > |{ r }_{ y|x }^{ * }| $, it suggests that y is more likely to be the "kernel cause" of x; equation (3) below implements the generalized correlation [\cite{Vinod2017}]; and
	\item (\textbf{M4}), causal additive modeling [\cite{Buhlmann2014}].
\end{itemize}

\begin{equation}
\left| \frac { \partial { g }_{ 1 }(Y|X) }{ \partial x }  \right| <\left| \frac { \partial { g }_{ 2 }(X|Y) }{ \partial y }  \right| 
\end{equation}

\noindent where ${ g }_{ 1 }$ is the model that predicts \textit{y}, and ${ g }_{ 2 }$ - the model that predicts \textit{x}.

\begin{equation}
|{Y}-pred({X})|=(|{ \epsilon  }_{ 1 }^{ \wedge  }|)<|{X}-pred({Y})|=(|{ \epsilon  }_{ 2 }^{ \wedge  }|)
\end{equation}

\noindent where \textit{pred(X)} is the value predicted by the model with respect to \textit{X}.

\begin{equation}
{ r }_{ y|x }^{ * }=sign({ r }_{ xy })\cdot \sqrt { 1-\frac { E{ (Y-E(Y|X) }^{ 2 } }{ var(Y) } }
\end{equation}

\noindent where $ { r }_{ xy } $ is the Pearson's correlation coefficient, \textit{var} is the variance and the expression inside the square root is the generalized measure of correlation (GMC) defined in [\cite{Zheng2012}].

The first three methods can be utilized with the \textbf{generalCorr} package [\cite{generalCorr}], and the last, CAM, using the \textbf{CAM} package [\cite{CAM}]. For clarity, no additional tuning and/or pruning was applied. 

We analyzed all pairs using all algorithms and stored the results with the provided ground truth. Then, we tried to assess the accuracies with confusion matrices. The combination of all results was tested with the majority voting technique (with the determination of the leader). Next, the accuracies of all combinations of results from the three algorithms were estimated in the same manner.

As the R package provided by Vinod enables the use of bootstrap analysis to determine the certainty of causal path direction, we assessed the quality of causal paths discovery in connection with the probability value, trying to find the best value, above which the highest accuracy of the causal direction indication is obtained, and below which the results should be questioned and probably not used for the final inference. Due to the fact that this operation is relatively time-consuming, and that some pairs have many observations, we decided to set the number of iterations to 10. We found that increasing the number to 20 or 50 does not change the final \textit{"p-cause"} value significantly.

Considering that causality is only sensible when at least a very small correlation is present and statistically significant (when the slope coefficient of the linear model is somehow important), we evaluated the accuracy when accepting only those cases that met two threshold conditions:

\begin{itemize}
	\item the absolute value of Pearson's correlation coefficient is greater than $0.1$, and
	\item the absolute value of the Bayesian correlation coefficient is greater than $0.1$ and $MFE>0.9$.
\end{itemize} 

Finally, we analyzed the impact of the number of observations in each pair on the accuracy by analyzing which number of observations maximizes balanced accuracy (the mean of sensitivity and specificity), overall accuracy and Cohen's Kappa.

All analysis was performed using R software, along with external packages [\cite{R}].

\section{Results}

The summary of accuracies obtained for separate algorithms is presented in Table 1. All of the considered approaches handle the data relatively poorly.

\tabcolsep=0.10cm
\begin{table}[!ht]
\caption{The summary of accuracies of causal path discoveries for all the considered algorithms, taken separately. The methods are described in the Materials and Methods section.}
\label{tab1}
\begin{center}
\begin{tabular}{l|rrrr}
\hline
\textbf{Method}     & \textbf{Accuracy} & \textbf{Sensitivity} & \textbf{Specificity} & \textbf{Cohen's Kappa} \\ \hline
\textbf{M1}                        & 0.4421                     & 0.4638                        & 0.3846                        & -0.1211                \\
\textbf{M2}                        & 0.6947                     & 0.7101                        & 0.6538                        & 0.3216                 \\
\textbf{M3}                        & 0.6316                     & 0.6087                        & 0.6923                        & 0.2452                 \\
\textbf{M4}                        & 0.5789                     & 0.5217                        & 0.7308                        & 0.1925                 \\ \hline
\end{tabular}
\end{center}
\end{table}
         
The results for the combination of four methods with each method as the leader, and for the best combination of three methods, is presented in Table 2. It appears that the best accuracy could be achieved with the \textbf{M2 + M3 + M4} criteria (with the final mark set by majority vote).

\tabcolsep=0.10cm
\begin{table}[!ht]
\caption{The summary of the accuracies of causal path discoveries taking into account all methods, with different leaders, along with the best combination of three methods. The methods are descrived in the Materials and Methods section.}
\label{tab2}
\begin{center}
\begin{tabular}{l|rrrr}
\hline
\textbf{Combination}     & \textbf{Accuracy} & \textbf{Sensitivity} & \textbf{Specificity} & \textbf{Cohen's Kappa} \\ \hline
\textbf{All} (with \textbf{M1} as leader)         & 0.6211                     & 0.6087                        & 0.6538                        & 0.2160                 \\
\textbf{All} (with \textbf{M2} as leader)         & 0.6947                     & 0.7101                        & 0.6538                        & 0.3216                 \\
\textbf{All} (with \textbf{M3} as leader)         & 0.6316                     & 0.5942                        & 0.7308                        & 0.2596                 \\
\textbf{All} (with \textbf{M4} as leader)         & 0.5789                     & 0.5217                        & 0.7308                        & 0.1925                 \\ 
\textbf{M2 + M3 + M4}                    & 0.8000                     & 0.8841                        & 0.5769                        & 0.4782                 \\ \hline
\end{tabular}
\end{center}
\end{table}

\begin{figure}[!ht]
\begin{center}
\includegraphics[width=4in]{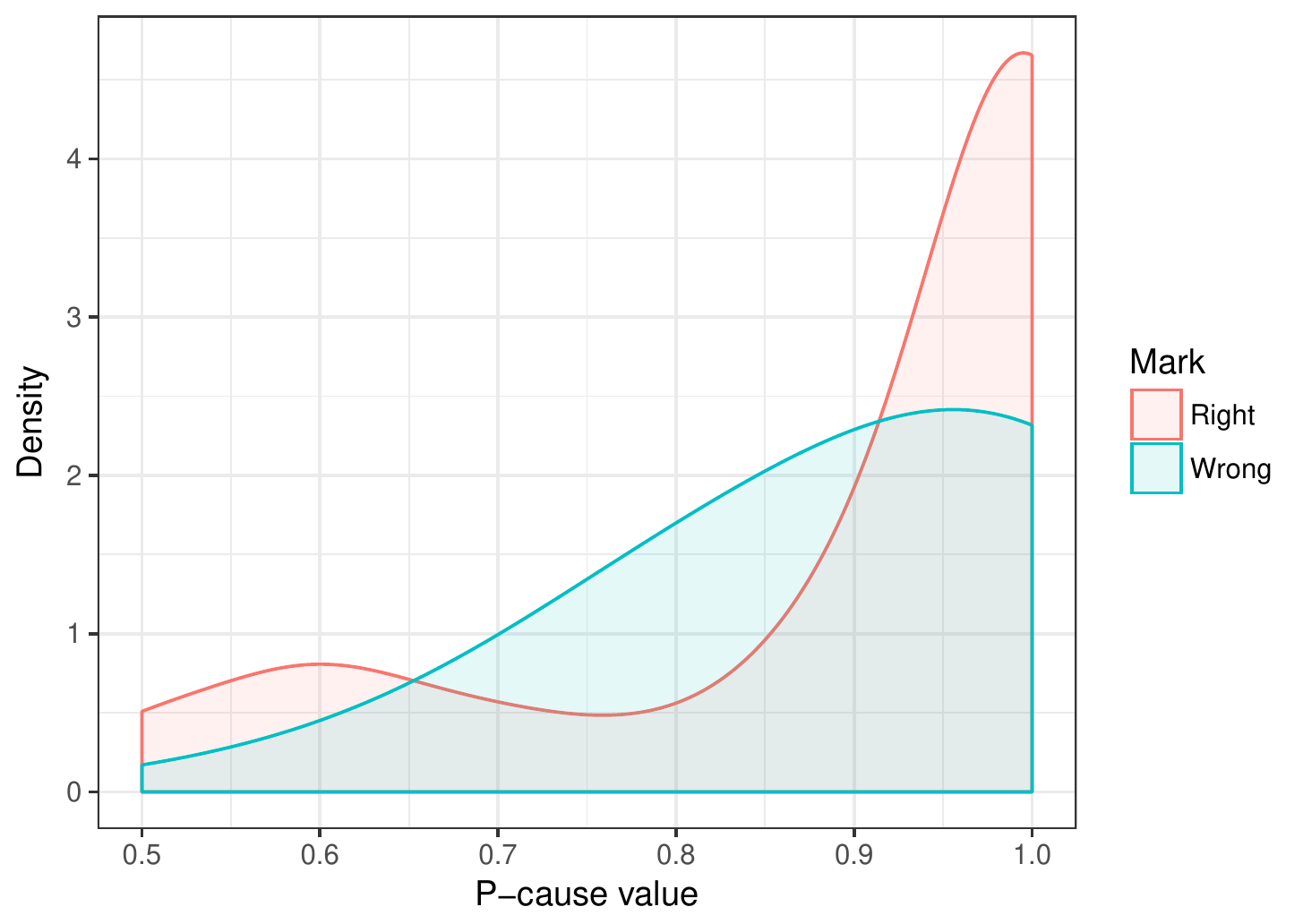}
\end{center}
\caption{Density distribution of correct indications in relation to bootstrap \textit{"p-cause"} values.} 
\label{fig1}
\end{figure}

A graph showing the distribution of correct indications in relation to \textit{"p-cause"} coming from bootstrap is presented in Figure \ref{fig1}. Through visual inspection, it seems that the level of \textit{"p-cause"}$=0.9$ is optimal for setting the heuristic threshold above which results are taken into account.

We analyzed the accuracies of the causal paths discoveries for the best combination of algorithms and for those pairs for which \textit{"p-cause"} was at least 0.9 (70 pairs; approximately $73\%$ of all results), and obtained an accuracy of about $83\%$, which is greater than that obtained without analyzing \textit{"p-cause"}. Interestingly, taking into account only those pairs for which \textit{"p-cause"}$=1$ does not increase the accuracy further, but only reduces the number of pairs included in the analysis (59 pairs; $62\%$). Taking only those cases where all methods produced the same result (unanimity; $55.8\%$ of results) actually reduces the accuracy. All statistics are provided in Table \ref{tab3}.

\tabcolsep=0.10cm
\begin{table}[!ht]
\caption{The summary of the accuracies of the causal path discoveries for the best combination, obtained for sets restricted with various criteria to increase the level of certainty.}
\label{tab3}
\begin{center}
\begin{tabular}{l|rrrr}
\hline
\textbf{Combination}     & \textbf{Accuracy} & \textbf{Sensitivity} & \textbf{Specificity} & \textbf{Cohen's Kappa} \\ \hline
\textbf{M2 + M3 + M4} (p\textgreater0.9) & 0.8286                     & 0.9020                        & 0.6316                        & 0.5518                 \\
\textbf{M2 + M3 + M4} (p=1)              & 0.8136                     & 0.9024                        & 0.6111                        & 0.5387                 \\
\textbf{M2 + M3 + M4} (unanimity)        & 0.7358                     & 0.7500                        & 0.7143                        & 0.4568        \\ \hline
\end{tabular}
\end{center}
\end{table}

\begin{figure}[!ht]
\begin{center}
\includegraphics[width=4in]{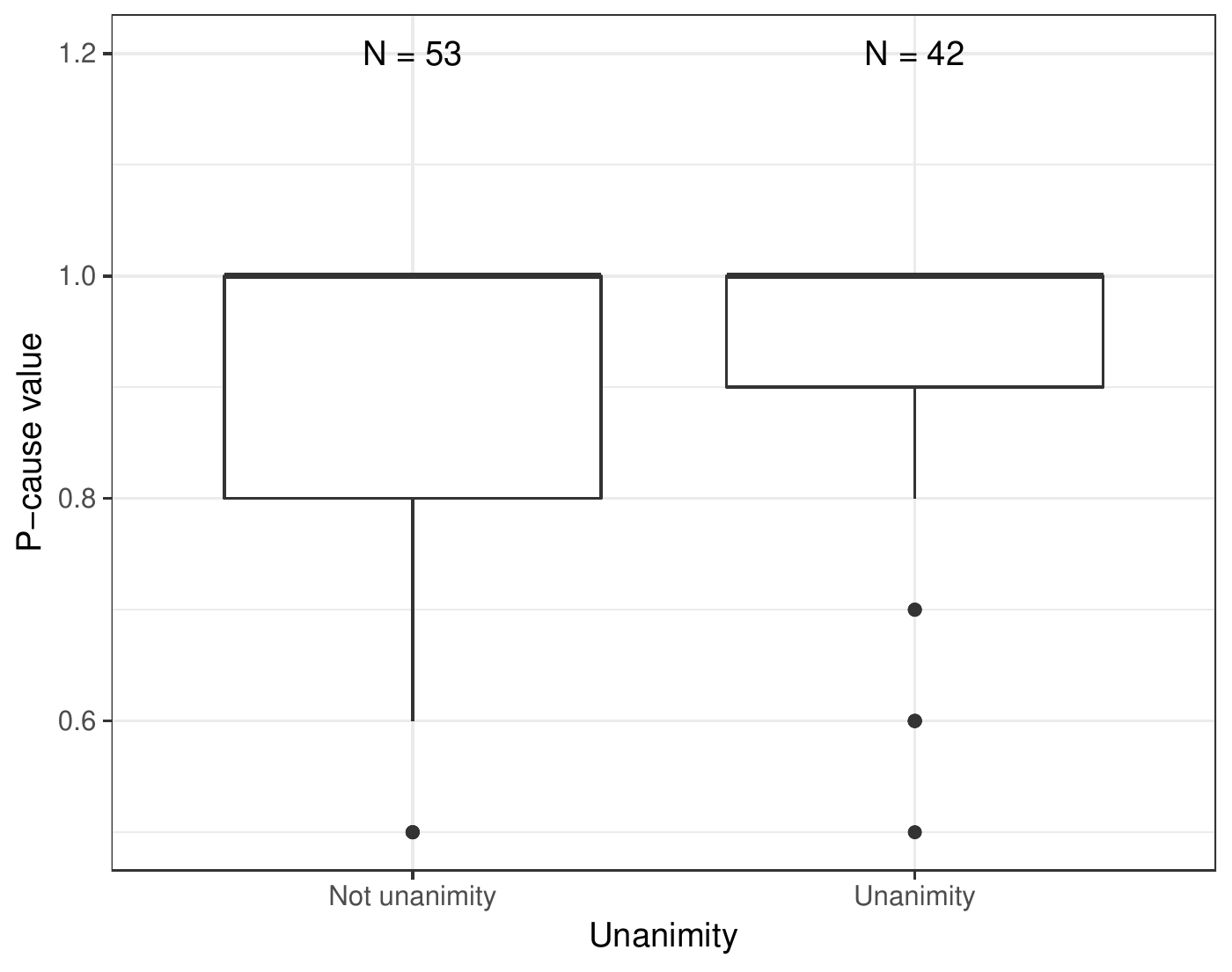}
\end{center}
\caption{Distribution of \textit{"p-cause"} in relation to unanimity of all three methods.} 
\label{fig2}
\end{figure}

The strength of the unanimity measure combined with the \textit{"p-cause"} parameter was assessed. A boxplot comparing the distribution of \textit{"p-cause"} values is presented in Figure \ref{fig2}. The results showed that the median value equals 1 for both cases. However, for "unanimity" cases, $75\%$ of results exceed 0.9, the level we chose to declare sensible causal path, which strengthens the inference possibilities.

Then, we analyzed (results in Table \ref{tab4}) the accuracy of causal path discovery for two sets of cases, with:

\begin{itemize}
	\item significant Pearson's correlation coefficients - $94.7\%$ pairs; interestingly, for all 4 pairs with insignificant correlation coefficients, the outcomes were correct;
	\item significant Bayesian correlation coefficients - $95.8\%$ pairs; the remaining 2 pairs with insignificant correlation coefficient also had correct results.
\end{itemize} 

\tabcolsep=0.10cm
\begin{table}[!ht]
\caption{The summary of the accuracies of the causal path discoveries for the best combination, after removing cases with insignificant correlation coefficients. PC crit. - the absolute value of Pearson's correlation coefficient is statistically significant and greater then $0.1$; BC crit. - the absolute value of the Bayesian correlation coefficient is statistically significant ($MFE>0.9$) and greater then $0.1$.}
\label{tab4}
\begin{center}
\begin{tabular}{l|rrrr}
\hline
\textbf{Combination}     & \textbf{Accuracy} & \textbf{Sensitivity} & \textbf{Specificity} & \textbf{Cohen's Kappa} \\ \hline
\textbf{M2 + M3 + M4} (PC crit.) & 0.7889                     & 0.8750                        & 0.5769                        & 0.4680                 \\
\textbf{M2 + M3 + M4} (BC crit.) &  0.7912                    & 0.8769                        & 0.5769                        & 0.4701                  \\ \hline
\end{tabular}
\end{center}
\end{table}

\begin{figure}[!ht]
\begin{center}
\includegraphics[width=3.5in]{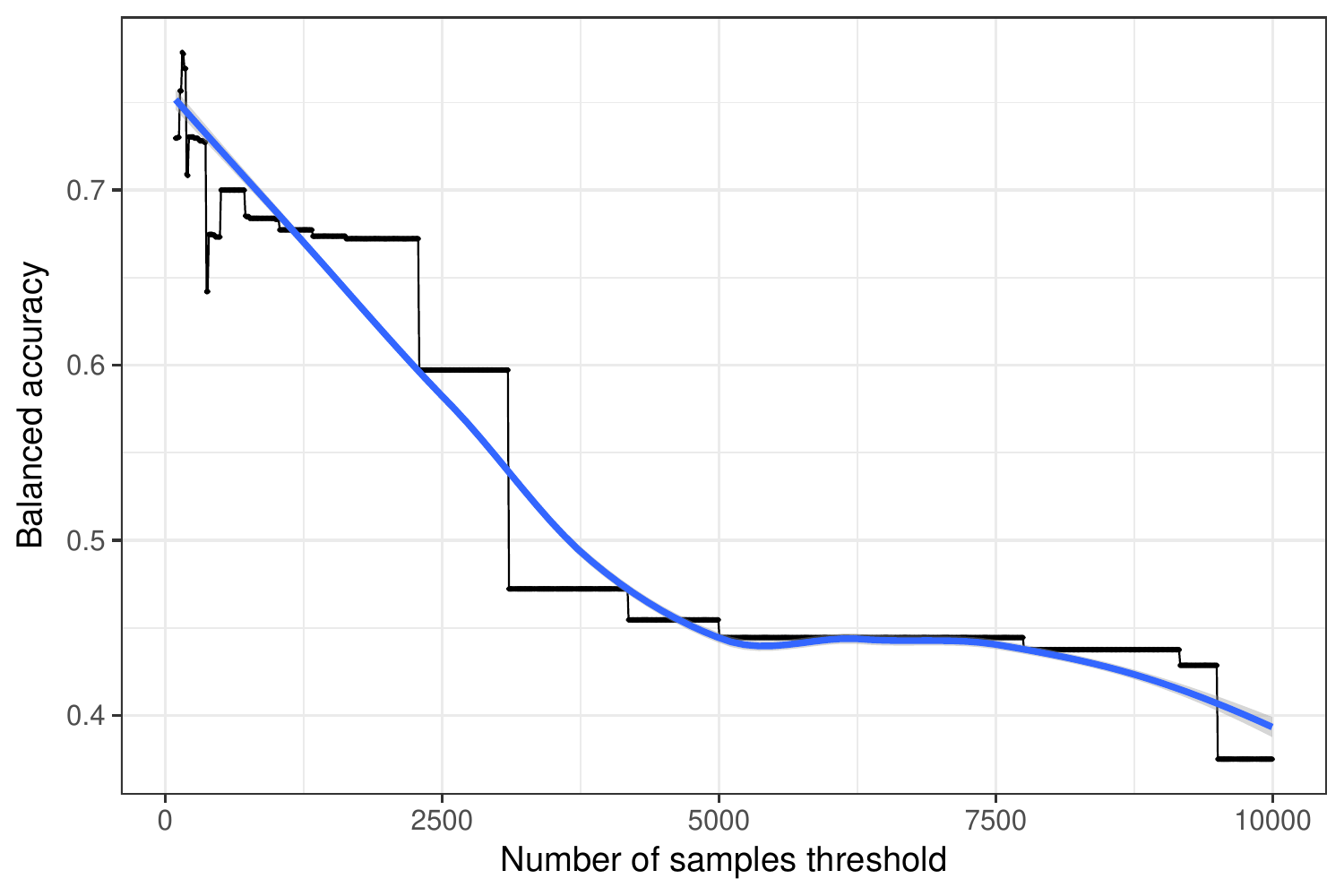}
\end{center}
\caption{Distribution of balanced accuracy in relation to the number of observations; the blue line represents the local polynomial regression fitting function.} 
\label{fig3}
\end{figure}

\begin{figure}[!ht]
\begin{center}
\includegraphics[width=3.5in]{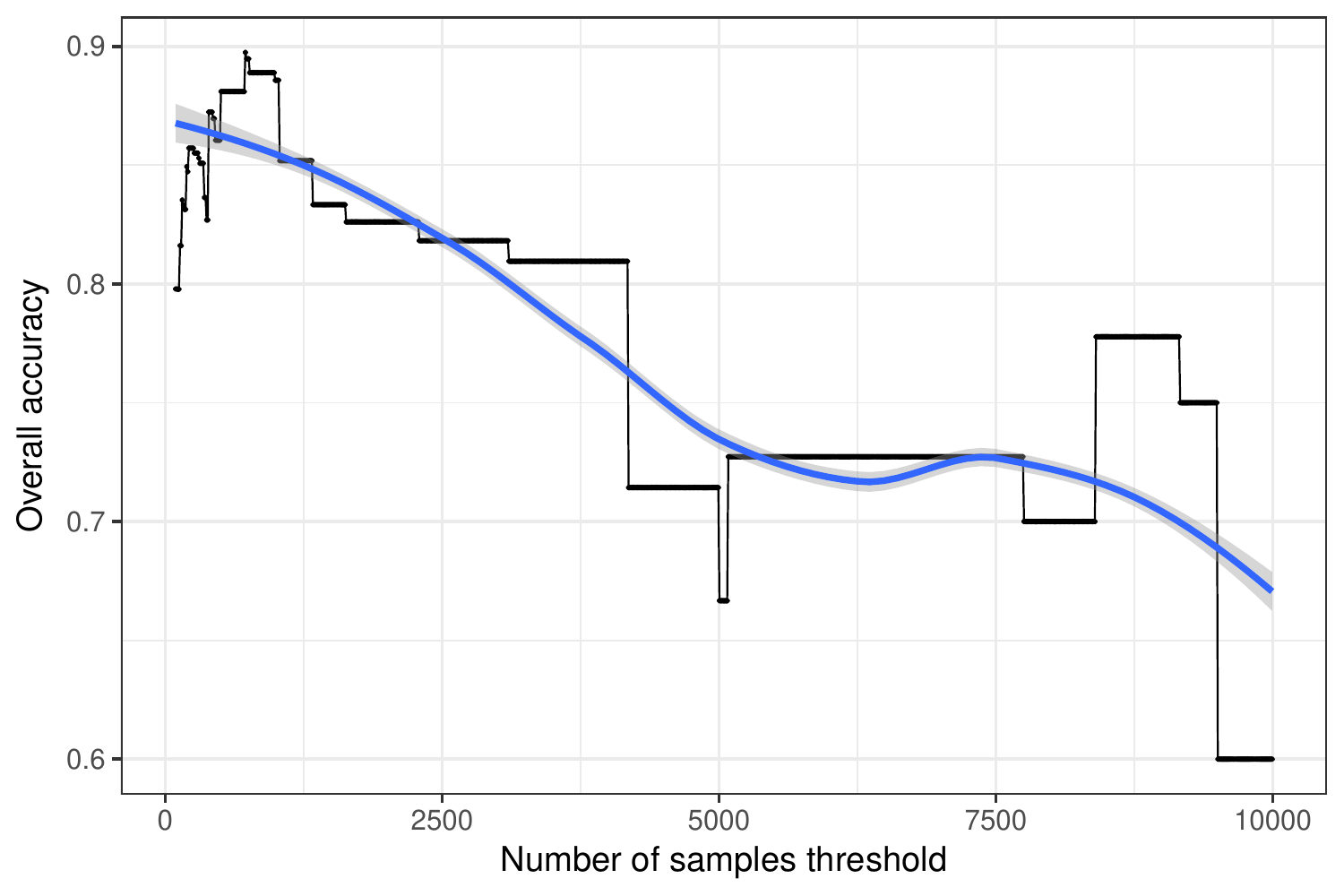}
\end{center}
\caption{Distribution of overall accuracy in relation to the number of observations; the blue line represents the local polynomial regression fitting function.} 
\label{fig4}
\end{figure}

\begin{figure}[!ht]
\begin{center}
\includegraphics[width=3.5in]{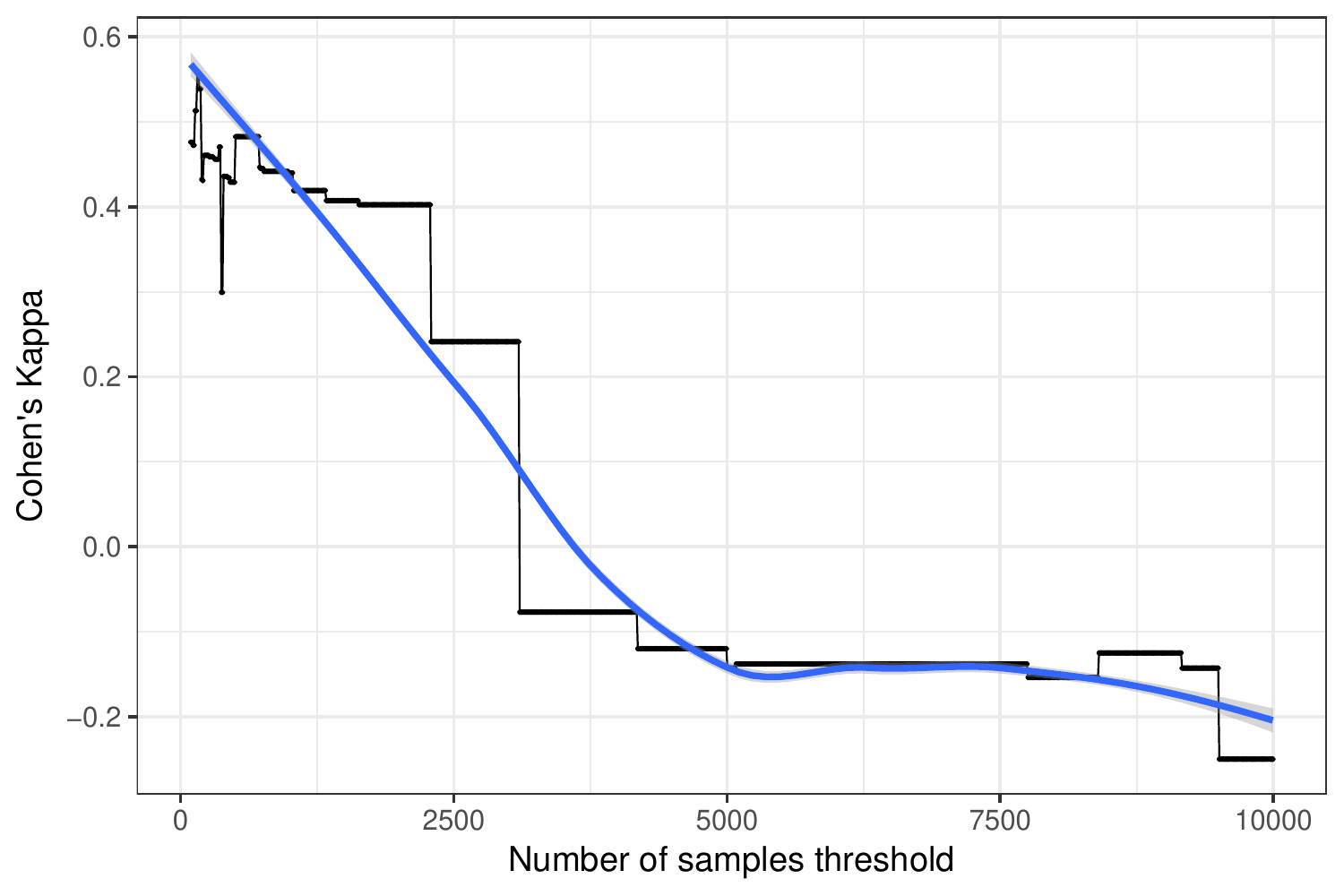}
\end{center}
\caption{Distribution of Cohen's Kappa in relation to the number of observations; the blue line represents the local polynomial regression fitting function.} 
\label{fig5}
\end{figure}

The curves of balanced accuracy (the mean of sensitivity and specificity), overall accuracy and Cohen's Kappa, in relation to the number of observations in the considered cases, are presented in Figures \ref{fig3}-\ref{fig5}, respectively. 

It appears that a small number of observations does not impact the accuracy; one might even say that a low number provides better accuracy. The maximum for balanced accuracy and Cohen's Kappa was found at 154, and at 724 for overall accuracy.

\section{Discussion}

In this paper, presented the results of a test of several methods intended for causal path discovery without prior knowledge of the data pairs coming from different fields of research. The best accuracy was obtained for the combination of generalized correlations framework, kernel regression absolute residuals criteria (AR) and causal additive modeling (CAM). All results were provided for the entire dataset cases; we did not have to divide the data into a training and test set (like in supervised learning approaches) because the analysis for each pair was treated as an independent step.

Several other methods could be included here based on meeting the basic criteria. For instance, mediation analysis and its extensions seem promising. However, it is relevant to problems of three or more variables, whereas the presented benchmark focused on the most elementary connections, within pairs. Still, the results from the analyzed framework could also serve as the input consideration for mediation analysis.

Otherwise, as in the case of the machine learning approach, we are not directly dependent on the training data, as well as on the consistency of the subsequent data used for testing and validation. 

In the contest report, it is stated that several baseline methods were utilized, e.g., additive-noise, latent variable or complexity-based models, as well as machine learning methods. Various aspects were evaluated, e.g., dependency, confounding, causality and final score. However, in order to compare them with the metrics used in this study, only causality seems relevant. Interestingly, such approaches gave slightly worse results, with about $76\%$ accuracy for real data as the best result. Several issues were also surveyed, e.g., preprocessing, feature extraction, dimensionality reduction, classification and time spent; this seems far more sophisticated than the approach presented in the benchmark study [\cite{Guyon2014}].

The presented approach demonstrates the possibility of establishing causal paths' direction based only on the data. We assume that this approach, when combined with information from experts, may enable universal assessment of causality. It seems impossible to choose real causal explanations only by looking at observed data: such prior knowledge would allow acquisition of causal information without performing interventions. In the study, the context is different. We believe that for causal analysis, especially in data cases for which bidirectional links are assumed as well as for small data sets, the results cannot be very dependent on prior knowledge, but they can complement it generally.

On the other hand, it seems that the presented approach could complement time-series analyses based on newer approaches and generalizations of Granger causality [\cite{Porta2016}]. We plan to connect such approaches and would also like to analyze the impact of many parameters, which can be established through the CAM process. 

\subsection{Limitations}

In the presented study, we did not analyze the distribution of data in each case, nor the cross-associations between pairs. This is because the data comes from various fields and such processing could not be coherent with the main objective. Still, the impact of mentioned aspects should be addressed. 

We wanted to focus on assessing elementary connections; as for the practical applications, the causal path for a set of more than two variables can be also established using the presented combination of methods, though this was not analyzed. 

Also, it turned out that it is reasonable to determine the direction of a relationship only for data that fulfills the bootstrap simulation criterion (\textit{"p-cause"}$>=0.9$); some results will be questioned, although their proportion seems sensible. Admittedly, as \textit{"p-cause"} comes from bootstrap simulation, its value is not very stable, so the proposed criterion should be viewed as as merely informative, not explicit.

\section{Conclusion}

In this benchmark study, an analysis of the accuracy of determining causal paths from the data from a causality contest was presented with the premise that the ground truth information is not taken into account for the discovery. The best overall accuracy was achieved for a combination of three methods: generalized correlations framework, kernel regression absolute residuals criteria (AR) and causal additive modeling (CAM). The most similar sensitivity and specificity values were obtained for AR criterion. 

We proposed to establish the causal path for about $73\%$ of the considered pairs based on the \textit{"p-cause"} criterion; the remaining cases are proposed to be left as indeterminate. The correlation coefficients and number of observations seemed to not affect the result much. 

In our opinion, the described approach can be used for preliminary dependence assessment, as an initial step for the commonly used causality assessment methods, or for comparison of data-driven findings with the ground truth. 

\lhead{}
\rhead{}
\bibliographystyle{chicago}

\newpage
\pagestyle{fancy}

\bigskip
\begin{center}
{\large\bf SUPPLEMENTARY MATERIAL}
\end{center}

\begin{description}

\item[T1:] The list of pairs from the database that were taken into consideration in the presented benchmark study [\cite{Mooij2016,Dua2017}]. The ground truth column stores the path of cause and effect; abbreviation: UCI - from UCI Machine Learning Repository.

\begin{center}
\begin{longtable}{p{0.06\linewidth}|p{0.46\linewidth}|p{0.48\linewidth}}
\hline
\textbf{No.}     & \textbf{Description} & \textbf{Ground truth (Path)}  \\ \hline
1 & Deutscher Wetterdienst Data & Altitude $\rightarrow$ Temperature \\  
2 & Deutscher Wetterdienst Data & Altitude $\rightarrow$ Precipitation \\  
3 & Deutscher Wetterdienst Data & Longitude $\rightarrow$ Temperature \\   
4 & Deutscher Wetterdienst Data & Altitude $\rightarrow$ Sunshine \\  
5 & Abalone data (UCI)  & Rings $\rightarrow$ Length \\   
6 & Abalone data (UCI) & Rings $\rightarrow$ Shell Weight  \\   
7 & Abalone data (UCI) & Rings $\rightarrow$ Diameter  \\   
8 & Abalone data (UCI) & Rings $\rightarrow$ Height  \\   
9 & Abalone data (UCI) & Rings $\rightarrow$ Whole Weight  \\   
10 & Abalone data (UCI) & Rings $\rightarrow$ Shucked Weight  \\  
11 & Abalone data (UCI) & Rings $\rightarrow$ Viscera Weight  \\  
12 & Census Income KDD (UCI) & Age $\rightarrow$ Wage per hour \\  
13 & Auto-Mpg Data (UCI) & Displacement $\rightarrow$ MPG \\  
14 & Auto-Mpg Data (UCI) & MPG $\rightarrow$ Horse power  \\  
15 & Auto-Mpg Data (UCI) & Weight $\rightarrow$ MPG  \\  
16 & Auto-Mpg Data (UCI) & Horse power $\rightarrow$ Acceleration  \\  
17 & Census Income KDD (UCI) & Age $\rightarrow$ Dividends from stock \\  
18 & Concentration of GG versus age for pediatrician & Age $\rightarrow$ GAG concentration \\  
19 & Old Faithful geyser data & Duration of eruption $\rightarrow$ Time to the next eruption \\  
20 & Deutscher Wetterdienst Data & Latitude $\rightarrow$ Temperature \\  
21 & Deutscher Wetterdienst Data & Longitude $\rightarrow$ Precipitation \\  
22 & Cardiac Arrhythmia Database (UCI) & Age $\rightarrow$ Height \\  
23 & Cardiac Arrhythmia Database (UCI) & Age $\rightarrow$ Weight \\  
24 & Cardiac Arrhythmia Database (UCI) & Age $\rightarrow$ Heart Rate  \\  
25 & Concrete Compressive Strength & Cement $\rightarrow$ Compressive Strength \\  
26 & Concrete Compressive Strength & Blast furnace slag $\rightarrow$ Compressive Strength \\  
27 & Concrete Compressive Strength & Fly Ash $\rightarrow$ Compressive Strength \\  
28 & Concrete Compressive Strength & Water $\rightarrow$ Compressive Strength  \\  
29 & Concrete Compressive Strength & Superplasticizer $\rightarrow$ Compressive Strength  \\  
30 & Concrete Compressive Strength & Coarse Aggregate $\rightarrow$ Compressive Strength  \\  
31 & Concrete Compressive Strength & Fine Aggregate $\rightarrow$ Compressive Strength  \\  
32 & Concrete Compressive Strength & Age $\rightarrow$ Compressive Strength  \\  
33 & BUPA liver disorders (UCI) & Drinks $\rightarrow$ Mean Corpuscular Volume \\  
34 & BUPA liver disorders (UCI) & Drinks $\rightarrow$ Alkaline Phosphotase  \\  
35 & BUPA liver disorders (UCI) & Drinks $\rightarrow$ Alanine Aminotransferase  \\  
36 & BUPA liver disorders (UCI) & Drinks $\rightarrow$ Aspartate Aminotransferase \\  
37 & BUPA liver disorders (UCI) & Drinks $\rightarrow$ Gamma-glutamyl Transpeptdase \\  
38 & Pima Indians Diabetes Database & Age $\rightarrow$ BMI \\  
39 & Pima Indians Diabetes Database & Age $\rightarrow$ 2-Hour serum insulin \\  
40 & Pima Indians Diabetes Database & Age $\rightarrow$ Diastolic Blood Pressure \\  
41 & Pima Indians Diabetes Database & Age $\rightarrow$ Plasma glucose concentration a 2 hours in an oral glucose tolerance test \\  
42 & Private archive of Bernward Janzing & Days of the year $\rightarrow$ Mean Daily Temperature of Furtwangen \\  
43 & Mean Daily Air temperature near surface & Day 50 $\rightarrow$ Day 51 \\  
44 & Mean Daily pressure at surface & Day 50 $\rightarrow$ Day 51 \\  
45 & Mean daily sea level pressure & Day 50 $\rightarrow$ Day 51 \\  
46 & Mean daily relative humidity near surface & Day 50 $\rightarrow$ Day 51 \\  
48 & Time series modelling of water resources and environmental systems & Outdoor temperature $\rightarrow$ Indoor temperature \\  
49 & Daily mean values of ozone and temperature of year 2009 in Lausanne-César-Roux & Temperature $\rightarrow$ Ozone \\  
50 & Daily mean values of ozone and temperature of year 2009 in Chaumont & Temperature $\rightarrow$ Ozone \\  
51 & Daily mean values of ozone and temperature of year 2009 in Davos-See & Temperature $\rightarrow$ Ozone \\  
56 & UNdata from {data.un.org} & Latitude of the Country's Capital $\rightarrow$ Fife Expectancy at Birth for female (2000-2005) \\  
57 & UNdata from {data.un.org} & Latitude of the Country's Capital $\rightarrow$ Fife Expectancy at Birth for female (1995-2000)  \\  
58 & UNdata from {data.un.org} & Latitude of the Country's Capital $\rightarrow$ Fife Expectancy at Birth for female (1990-1995)  \\  
59 & UNdata from {data.un.org} & Latitude of the Country's Capital $\rightarrow$ Fife Expectancy at Birth for female (1985-1990)  \\  
60 & UNdata from {data.un.org} & Latitude of the Country's Capital $\rightarrow$ Fife Expectancy at Birth for male (2000-2005) \\  
61 & UNdata from {data.un.org} & Latitude of the Country's Capital $\rightarrow$ Fife Expectancy at Birth for male (1995-2000)  \\  
62 & UNdata from {data.un.org} & Latitude of the Country's Capital $\rightarrow$ Fife Expectancy at Birth for male (1990-1995)  \\  
63 & UNdata from {data.un.org} & Latitude of the Country's Capital $\rightarrow$ Fife Expectancy at Birth for male (1985-1990)  \\  
64 & UNdata from {data.un.org} & Population with sustainable access to improved drinking water sources $\rightarrow$ Infant mortality rate \\  
65 & Financial data from Jan. 4, 2000 to Jun. 17, 2005 & Stock returns of Hang Seng Bank $\rightarrow$ Stock return of HSBC Hldgs \\  
66 & Financial data from Jan. 4, 2000 to Jun. 17, 2005 & Stock returns of Hutchison $\rightarrow$ Stock return of Cheung kong  \\  
67 & Financial data from Jan. 4, 2000 to Jun. 17, 2005 & Stock returns of Cheung kong $\rightarrow$ Stock return of Sun Hung Kai Prop.  \\  
68 & Internet connections and traffic at the MPI for Intelligent Systems & Open HTTP Connections $\rightarrow$ Bytes Sent \\  
69 & Temperature data provided by Joris M. Mooij & Outdoor temperature $\rightarrow$ Indoor temperature \\  
72 & Sunspot data & Sunspot Area $\rightarrow$ Global Mean Temperature Anomalies \\  
73 & Energy - emission data from 152 countries between 1960 and 2005 & Energy use $\rightarrow$ CO2 emissions  \\  
74 & UNdata from {data.un.org} & Gross National Income $\rightarrow$ Life Expectancy at Birth  \\  
75 & UNdata from {data.un.org} & Gross National Income $\rightarrow$ Under 5 Mortality Rate\\  
76 & Data for 174 countries provided by Food and Agriculture Organization of the United Nations & Average Annual Rate of Change of Population $\rightarrow$ Average Annual Rate of Change of Total Dietary Consumption for Total Population \\  
77 & Data from 1985 to 2008, provided by Bernward Janzing & Solar Radiation measured in Furtwangen $\rightarrow$ Daily Average Temperature \\  
78 & Light Response Data & Photosynthetic Photon Flux Density $\rightarrow$ Net Ecosystem Productivity \\  
79 & Light Response Data & Photosynthetic Photon Flux Density, diffusive $\rightarrow$ Net Ecosystem Productivity  \\  
80 & Light Response Data & Photosynthetic Photon Flux Density, direct $\rightarrow$ Net Ecosystem Productivity  \\  
84 & Data for 3102 counties in US in 1980 & Natural Logarithm of the Corresponding Population $\rightarrow$ Natural Logarithm of Employment \\  
85 & Milk protein trial data used by Verbyla and Cullis (1990) & Time to Take Weekly Measurements (from 1 to 14) $\rightarrow$ Protein Content of the Milk Produced by each Cow \\  
87 & Whistler Daily Snowfall & Mean Temperature $\rightarrow$ Total Snow \\  
88 & "bone" data set from CRAN & Age $\rightarrow$ Relative Spinal Bone Mineral Density \\  
89 & Data taken from Solly et al. (2014) on decomposition rates in forests and grasslands & Mass Loss in forests after 6 months $\rightarrow$ Mass Loss in forests after 1 year \\  
90 & Data taken from Solly et al. (2014) on decomposition rates in forests and grasslands & Mass Loss in grasslands after 6 months $\rightarrow$ Mass Loss in grasslands after 1 year  \\  
91 & Data taken from Solly et al. (2014) on decomposition rates in forests and grasslands & Clay content in soil $\rightarrow$ Soil moisture at 10cm depth \\  
92 & Data taken from Solly et al. (2014) on decomposition rates in forests and grasslands & Clay content in soil $\rightarrow$ Organic C Content in Soil \\  
93 & MOPEX data set over 1948 to 2004 & Average Precipitation $\rightarrow$ average Runoff \\  
94 & Data from a regional energy distributor in Turkey & Hour of the day $\rightarrow$ Temperature \\  
95 & Data from a regional energy distributor in Turkey & Hour of the day $\rightarrow$ Total Electricity Consumption \\  
96 & Data from a regional energy distributor in Turkey & Temperature $\rightarrow$ Total Electricity Consumption  \\  
97 & Data on speed of a ball on a ball track for children, recorded by Dominik Janzing & Initial speed $\rightarrow$ Speed at later position \\  
98 & Data on speed of a ball on a ball track for children, recorded by Dominik Janzing & Initial speed $\rightarrow$ Final speed \\  
99 & 'nlschools' from the R MASS package & Social-Economic Status of Pupil's Family $\rightarrow$ Language Test Score \\  
100 & 'cpus' from the R MASS package & Cycle time $\rightarrow$ Published performance on a benchmark mix \\ 
101 & Brightness of screen & Grey value of a pixel $\rightarrow$ Light intensity seen by a photo diode \\  
102 & Data on speed of a ball, recorded by Dominik Janzing & Position on the ball track where the ball starts $\rightarrow$ Time interval between passing the first and the second light barrier \\ 
103 & Data on speed of a ball, recorded by Dominik Janzing & Position on the ball track where the ball starts $\rightarrow$ Time interval between passing the third and the fourth light barrier  \\ 
104 & Data on speed of a ball, recorded by Dominik Janzing & Time interval between passing the first and the second light barrier $\rightarrow$ Time interval between passing the third and the fourth light barrier  \\ 
106 & Speed of an electric toy locomotive & Electric voltage $\rightarrow$ Time required for passing one round \\  
108 & Data on heat bath of a Striling engine & Temperature $\rightarrow$ Time for 1/6 rotation \\ \hline
\end{longtable}
\end{center}

\end{description}

\end{document}